\documentclass[twocolumn,showpacs,amsmath,amssymb,prl]{revtex4}
\usepackage{amsxtra}
\usepackage{amssymb}
\usepackage{amsmath}
\usepackage{graphicx}

\newcommand {\rhovec}{\ensuremath \boldsymbol{\rho}}

\begin{document}
\title{Ghost Imaging: What is quantum, what is not}
\date{\today}
\author{Baris I. Erkmen}
\email{erkmen@mit.edu}
\author{Jeffrey H. Shapiro}
\affiliation{Massachusetts Institute of Technology, Research Laboratory of Electronics, Cambridge, Massachusetts 02139, USA}

\begin{abstract}
We provide a unified treatment of classical and quantum Gaussian-state sources that unambiguously identifies which features of ghost imaging are strictly quantum mechanical. We show that ghost-image formation is fundamentally classical, with the image being expressible in terms of the phase-insensitive and phase-sensitive cross correlations between the detected fields. We then consider ghost-imaging scenarios with either phase-insensitive or phase-sensitive sources, where the former are always classical but the latter may be classical or quantum mechanical.  We show that if their auto-correlations are identical, then a quantum source provides resolution improvement in its near-field and field-of-view improvement in its far field when compared to a classical source.  
\end{abstract}
\pacs{42.30.Va, 42.50.Ar, 42.50.Dv}
\maketitle 
Ghost imaging is the acquisition of an object's transmittance pattern by means of intensity correlation measurements. Its first demonstration utilized a biphoton source, thus the image was interpreted as a quantum phenomenon owing to the entanglement of the source photons \cite{Pittman}. However, subsequent experimental \cite{Bennink,Valencia,Ferri} and theoretical \cite{Gatti:three,Gatti} considerations have demonstrated that ghost imaging can be performed with thermalized laser light. Whereas the biphoton requires a quantum description for its photodetection statistics, thermal light does not, i.e., it can be regarded as a classical electromagnetic wave whose photodetection statistics can be treated via shot-noise theory.  This disparity has sparked interest in determining whether the fundamental physics of ghost imaging require a non-classical source \cite{Bennink:two,DAngelo,Cai,Scarcelli}. In this paper, we shall treat ghost imaging with classical and non-classical Gaussian-state sources to quantify the classical/quantum boundary. We begin by deriving the image---in terms of the detected fields' cross-correlation functions---for arbitrary two-mode, coherence separable Gaussian-state sources. We then compare image resolution and field-of-view for near-field and far-field ghost imaging using three types of sources: thermal light, classical phase-sensitive light and quantum phase-sensitive light. We conclude with a discussion of the relevant physics in ghost imaging, emphasizing which aspects are classical and which are quantum.

\begin{figure}[t]
\begin{center}
\includegraphics[width= 3in]{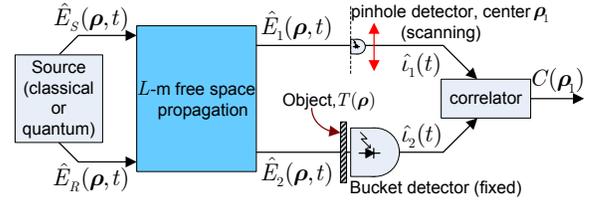}
\end{center}
\caption{(Color online) A simple ghost imaging setup.} \label{GI:propagation}
\end{figure}

Consider the ghost imaging setup shown in Fig.~\ref{GI:propagation}. Here, $\hat{E}_{S}(\rhovec,t)e^{-i \omega_{0}t}$ and $\hat{E}_{R}(\rhovec,t)e^{-i \omega_{0}t}$ are a pair of scalar, positive frequency, $z$-propagating field operators normalized to photon-units, where $\omega_{0}$ is their common center frequency, $\rhovec$ is the transverse coordinate with respect to each one's optical axis, and
\begin{equation}
[\hat{E}_{m}(\rhovec_{1},t_{1}),\hat{E}_{m}^{\dagger}(\rhovec_{2},t_{2})]=\delta(\rhovec_{1} - \rhovec_{2}) \delta(t_{1} - t_{2})\,,
\end{equation}
for $m \in \{S,R\}$, are the non-zero commutators. Both beams undergo quasimonochromatic paraxial diffraction over an $L$-m-long free-space path, yielding measurement-plane field operators
\begin{equation}
\hat{E}_{\ell}(\rhovec,t) = \int\! d \rhovec'\, \hat{E}_{m}\big(\rhovec',t-L/c \big) \frac{k_{0} e^{i k_{0} |\rhovec-\rhovec'|^2/2L}}{i 2 \pi L}, \label{FS:prop}
\end{equation} 
where $(\ell,m) = \{(1,S), (2,R)\}$, $c$ is the speed of light and $k_{0} = \omega_{0}/c$. The first field, $\hat{E}_{1}(\rhovec,t)$, illuminates a quantum-limited pinhole photodetector located at transverse coordinate $\rhovec_{1}$. The second field, $\hat{E}_{2}(\rhovec,t)$, illuminates an amplitude-transmission mask $T(\rhovec)$, located immediately in front of a quantum-limited bucket photodetector with sensitive region $\rhovec \in \mathcal{A}_{2}$. The product of the photocurrents from these sub-unity quantum efficiency, finite-bandwidth detectors is time averaged to estimate their ensemble-average cross correlation, $C(\rhovec_{1})$.  This process is repeated, as $\rhovec_{1}$ is scanned over the plane, to obtain the ghost image of the object's intensity transmission $|T(\rhovec)|^2$.

Without appreciable loss of generality, we shall assume that $\hat{E}_{S}$ and $\hat{E}_{R}$ are in a zero-mean, space-time coherence separable, jointly-Gaussian quantum state.   Such a state is completely characterized by the field operators' phase-insensitive (normally-ordered) and phase-sensitive auto- and cross correlations. We allow the source fields to have both phase-insensitive and phase-sensitive cross correlations, but for simplicity we shall assume they carry no phase-sensitive auto-correlations. Because \eqref{FS:prop} is linear, $\hat{E}_{1}$ and $\hat{E}_{2}$ are also in a zero-mean, coherence separable, jointly-Gaussian state, and hence fully determined by 
\begin{align}
\langle \hat{E}_{m}^{\dagger}(\rhovec_1,t_1) \hat{E}_{\ell}(\rhovec_2,t_2) \rangle &= K^{(n)}_{m,\ell}(\rhovec_1, \rhovec_2) R^{(n)}_{m,\ell}(t_2-t_1) \label{CORR:PIS}\\ 
\langle \hat{E}_{1} (\rhovec_1,t_1) \hat{E}_{2}(\rhovec_2,t_2) \rangle  &= K^{(p)}_{1,2}(\rhovec_1, \rhovec_2) R^{(p)}_{1,2} (t_2-t_1), \label{CORR:PS}
\end{align}
for $m,\ell \in \{1,2\}$. The moment-factoring property of Gaussian states allow us to write the photocurrent cross-correlation at each $\rhovec_{1}$ as follows,
\begin{eqnarray}
C(\rhovec_{1}) &=& C_{0} + C_{n} \int_{\mathcal{A}_{2}}\! d\rhovec\,  |K^{(n)}_{1,2}(\rhovec_1, \rhovec)|^{2}  |T(\rhovec)|^{2} \nonumber \\[.05in] &+& C_{p}  \int_{\mathcal{A}_{2}}\! d\rhovec\,  |K^{(p)}_{1,2}(\rhovec_1, \rhovec)|^{2}  |T(\rhovec)|^{2}\,, \label{GI:corr}
\end{eqnarray}
where $C_{0}$, $C_{n}$ and $C_{p}$ are positive constants, with the former constituting a non-image-bearing background proportional to the product of the incident photon fluxes, and the latter two depending on the temporal cross correlations $R^{(n)}_{1,2}(\tau)$ and $R^{(p)}_{1,2}(\tau)$, respectively. The ghost image is therefore a linear, space-varying filtered version of the object's intensity transmission, whose filter response depends on the spatial cross-correlations between $\hat{E}_1$ and $\hat{E}_2$.  Because any pair of phase-insensitive and phase-sensitive cross correlations may be associated with a classical Gaussian state (i.e., a Gaussian state with a proper $P$-representation), ghost-image formation is intrinsically classical \cite{classical}.

Suppose the signal and reference fields have the same Gaussian-Schell model phase-insensitive auto-correlation 
\begin{eqnarray}
\lefteqn{K^{(n)}_{m,m}(\rhovec_1, \rhovec_2) R^{(n)}_{m,m}(t_2-t_1)=} \nonumber\\[.12in]
&&\frac{2 P}{\pi a_{0}^2} e^{-(|\rhovec_{1}|^2+ |\rhovec_{2}|^2)/a_{0}^2 - |\rhovec_{2} - \rhovec_{1}|^2/2 \rho_{0}^2 } e^{-(t_2 - t_1)^2/2T_{0}^{2}} \label{GSM:PIS}\,,
\end{eqnarray}
where $a_{0}$ denotes the beam radius, $\rho_{0}$ is the coherence length (assumed to satisfy the low-coherence condition $\rho_{0} \ll a_{0}$), and $T_{0}$ is the coherence time of the $P$-photons/sec signal ($m= S$) and reference ($m=R$) fields. Two-mode Gaussian state sources with no phase-sensitive cross correlation between the signal and reference modes (i.e.,  $K_{S,R}^{(p)} = 0$) always have a proper $P$-representation, and hence are classical. A well-known example is thermalized laser light, generated by passing a continuous-wave laser beam through a rotating ground-glass diffuser followed by a 50/50 beam splitter. These two beams have maximum $|\langle \hat{E}_S^\dagger(\rhovec_1,t_1)\hat{E}_R(\rhovec_2,t_2)\rangle|$, given by \eqref{GSM:PIS}. Taking the phase of this cross correlation to be zero we get
\begin{eqnarray}
C(\rhovec_{1}) &=& C_0 + C_{n} (2 P/\pi a_{0}^2 )^{2} \, e^{-2 |\rhovec_{1}|^2/a^{2}_{0}} 
\nonumber \\[.05in] &\times&
\int_{\mathcal{A}_{2}}\! {\rm d}\rhovec \, e^{-|\rhovec_{1} - \rhovec|^2/\rho_{0}^{2}} e^{-2 |\rhovec|^2/a^{2}_{0}} |T(\rhovec)|^{2}
\label{Thermal:NF}
\end{eqnarray}
in the near field of the source (i.e., $L$ such that the Fresnel number product $D_{0} \equiv k_{0} a_{0} \rho_{0}/ 2 L \gg 1$). From \eqref{Thermal:NF} we see that the phase-insensitive cross correlation between $\hat{E}_{1}$ and $\hat{E}_{2}$ has three important consequences. First, the ghost image is space-limited by the average intensity profile of $\hat{E}_{2}$, which affirms that the object must be placed in the field of view $a_{0}$ of the reference beam \cite{FOV}. Second, the finite size of $\hat{E}_{1}$ restricts the useful transverse scanning range of the pinhole detector to a radius $a_{0}$. Finally, and most importantly, the finite cross-correlation coherence length $\rho_{0}$ limits the resolution of the image.  When space-limiting and finite detector area can be neglected, the ghost image in \eqref{Thermal:NF} is proportional to the convolution of the object's intensity transmission, $|T(\rhovec)|^2$, with the Gaussian point-spread function $e^{-|\rhovec|^2/\rho_0^2}$.  Thus the resolution---radius to the $e^{-2}$-level in the point-spread function---is $\sqrt{2}\rho_0$.

If the object is in the source's far-field ($D_{0} \ll 1$), the image still has the form \eqref{Thermal:NF}, but the beam diameter and the coherence length diffract in equal proportion, so that $a_{0}$ and $\rho_{0}$ must be replaced with $a_{L} = 2 L / k_{0} \rho_{0}$ and $\rho_{L} = 2 L / k_{0} a_{0}$ respectively \cite{Mandel}. Therefore, the far-field field-of-view increases to $a_{L}$ while the image resolution degrades to $2\sqrt{2} L / k_{0} a_{0}$.  Thus, so long as field-of-view is not the limiting factor, it is more desirable to place the object in the near field of the source. When the object is far from the source, the source coherence can be transferred to the object plane with lenses.

When Gaussian-state signal and reference fields have a non-zero phase-sensitive cross correlation, their joint state need \em not\/\rm\ have a proper $P$-representation, viz., the state may be non-classical. Consider signal and reference fields with Schell-model phase-insensitive auto-correlations given by 
 $A^{*}(\mathbf{x}_{1}) A(\mathbf{x}_{2}) g^{(n)}(\mathbf{x}_{2}-\mathbf{x}_{1})$, for $\mathbf{x} \equiv (\rhovec, t)$, a phase-sensitive cross correlation of similar form, $A(\mathbf{x}_{1}) A(\mathbf{x}_{2}) g^{(p)}(\mathbf{x}_{2}-\mathbf{x}_{1})$, and no phase-insensitive cross correlation. Here we shall require $|A(\mathbf{x})| \leq 1$, so that this function may be regarded as a (possibly complex-valued) spatio-temporal attenuation of two homogenous and stationary random fields with phase-insensitive auto-correlations $g^{(n)}(\mathbf{x}_{2}-\mathbf{x}_{1})$, and a phase-sensitive cross correlation $g^{(p)}(\mathbf{x}_{2}-\mathbf{x}_{1})$. Then, the phase-sensitive cross correlation spectrum, given by the 3D Fourier transform $\tilde{g}^{(p)}(\mathbf{f}) \equiv \mathcal{F}\{ g^{(p)}(\mathbf{x}) \}$, must satisfy
\begin{equation}
|\tilde{g}^{(p)}(\mathbf{f})| \leq \sqrt{\tilde{g}^{(n)}(\mathbf{f}) (1 + \tilde{g}^{(n)}(\mathbf{f}))}\,, \label{NCL:PHS}
\end{equation}
where $\tilde{g}^{(n)}(\mathbf{f}) \equiv \mathcal{F}\{ g^{(n)}(\mathbf{x}) \}$ is an even phase-insensitive auto-correlation spectrum. However, a \em classical\/\rm\ Gaussian state has its $\tilde{g}^{(p)}(\mathbf{f})$ limited by the more restrictive necessary and sufficient condition \cite{Shapiro:Gaussian}
\begin{equation}
|\tilde{g}^{(p)}(\mathbf{f})|  \leq \tilde{g}^{(n)}(\mathbf{f})\,. \label{CL:PHS}
\end{equation}
Applying \eqref{CL:PHS} to the Gaussian-Schell model auto-cor\-relations in \eqref{GSM:PIS} and taking the phase to be zero, we find that the maximum classical $\langle \hat{E}_S(\rhovec_1,t_1)\hat{E}_R(\rhovec_2,t_2)\rangle$ is also equal to \eqref{GSM:PIS}. 
The photocurrent correlation $C(\rhovec_{1})$ in the source's near field is found to coincide with the thermal light case in \eqref{Thermal:NF}, yielding the same field-of-view and resolution. However, phase-sensitive coherence propagates differently than phase-insensitive coherence \cite{ErkmenShapiro:CohThy}. In particular, deep in the far-field---when $D_{0}\ll 1$ and $k_{0} a_{0}^2 / 2 L \ll 1$---we get,
\begin{eqnarray}
C(\rhovec_{1}) &=& C_{0} + C_{p} (2 P/\pi a_{L}^{2})^2 \, e^{- 2 |\rhovec_{1}|^2/a^{2}_{L}} \nonumber \\[.05in] &\times&
\int_{\mathcal{A}_{2}} \!d\rhovec \, e^{-|\rhovec_{1} + \rhovec|^2/\rho_{L}^{2}} e^{-2 |\rhovec|^2/a^{2}_{L}} |T(\rhovec)|^{2}\,,
\label{PHS_CL:FF}
\end{eqnarray}
where  $a_{L} = 2 L / k_{0} \rho_{0}$ and $\rho_{L} = 2 L / k_{0} a_{0}$ are the same as the thermal light case. Therefore, \eqref{PHS_CL:FF} has the same resolution and field-of-view as the far- field image from thermal light, but the image is now inverted, i.e., the ghost image is proportional to the convolution of $|T(-\rhovec)|^2$ with $e^{-|\rhovec|^2/\rho^{2}_{L}}$.

The maximum non-classical phase-sensitive cross correlation satisfies \eqref{NCL:PHS} with equality. We will restrict our attention to two limiting cases in which such a phase-sensitive cross correlation is coherence separable, so that we may utilize the machinery developed earlier in this paper. If $\tilde{g}^{(n)}(\mathbf{f}) \gg 1$, the distinction between quantum and classical sources diminish and the classical phase-sensitive results apply. On the other hand, if $\tilde{g}^{(n)}(\mathbf{f}) \ll 1 $, then $|\tilde{g}^{(p)}(\mathbf{f})|\approx \sqrt{\tilde{g}^{(n)}(\mathbf{f})}$, and the source exhibits strongly non-classical behavior. Assuming a real $\tilde{g}^{(p)}(\mathbf{f})$ and using \eqref{GSM:PIS}, this approximation holds when $P T_{0} \rho_{0}^{2} / a_{0}^2 \ll 1$, yielding 
\begin{eqnarray}
\lefteqn{\langle \hat{E}_S(\rhovec_1,t_1)\hat{E}_R(\rhovec_2,t_2)\rangle \propto }\nonumber \\[.12in]
&&
e^{ -(|\rhovec_{1}|^2+ |\rhovec_{2}|^2)/a_{0}^{2} - |\rhovec_{2} - \rhovec_{1}|^2/\rho_{0}^{2} } e^{-(t_2 - t_1)^2/T_{0}^{2}}\,. \label{GSM:PS_NCL}
\end{eqnarray} 
Equation~\eqref{GSM:PS_NCL} is a Gaussian-Schell model correlation, so
\begin{eqnarray}
\lefteqn{C(\rhovec_{1}) = C_0 + C_p'e^{-2|\rhovec_1|^2/a_0^2}} \nonumber \\[.12in] 
&\times & \int_{\mathcal{A}_{2}}\! d\rhovec \, 
e^{-2 |\rhovec_{1} - \rhovec|^2/\rho_{0}^{2}} e^{-2|\rhovec|^2/a_0^2} |T(\rhovec)|^{2}
\label{PHS:NF_NCL}
\end{eqnarray}
in the source's near field, where $C_p'$ is a positive constant.  
Here the ghost image has field-of-view $a_{0}$ and resolution $\rho_{0}$, which is a factor-of-$\sqrt{2}$ better than the resolution from (phase-sensitive or phase-insensitive) classical sources with identical source auto-correlations.  

The far-field $C(\rhovec_{1})$ for the non-classical source is given by \eqref{PHS_CL:FF}, with $\rho_{L} =  2 L / k_{0} a_{0}$, $a_{L} = 2\sqrt{2} L /k_{0} \rho_{0}$, and $C_p'$ replacing $C_p$. Hence the far-field resolution for the quantum source equals that of the  classical sources considered earlier, but the field of view $a_{L}$, is $\sqrt{2}$-larger. It is worth pointing out that these enhancements are due to the broadening of the weak spectrum, $\tilde{g}^{(n)}(\mathbf{f})$, when its square-root is taken. Thus the enhancement factor depends on the shape of the cross-correlation spectrum.
\begin{figure}[t]
\begin{center}
\includegraphics[width= 3in]{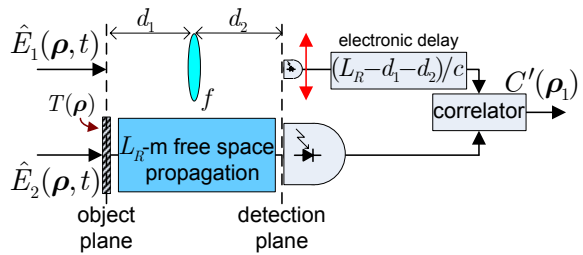}
\end{center}
\caption{(Color online) Ghost imaging setup with relay optics.} \label{GI:detectorplane}
\end{figure}
\section{Discussion}

The underlying principle in ghost imaging, whether the source is classical or quantum, is the non-zero intensity cross-correlation between the signal and reference beams. For Gaussian-state sources, this intensity cross correlation is given in terms of the second-order correlation functions, and both  phase-sensitive and phase-insensitive cross correlations can contribute to the image. Any pair of phase-sensitive and phase-insensitive cross-correlation functions can be obtained with two classical Gaussian-state fields, so long as there are no restrictions on these fields' auto-correlation functions. In this respect, the ghost image does \em not\/\rm\ contain any quantum signature \em per se\/\rm. However, if we compare sources that have identical auto-correlation functions, we find that non-classical fields with low brightness and maximum phase-sensitive cross correlation offer a spatial resolution advantage in the source's near field and a field-of-view extension in its far field. The field-of-view in the near field and the resolution in the far-field are determined by the beam sizes at the source, and hence are identical for classical and non-classical fields. 

It is worth addressing the physical implications of the $P T_{0} \rho_{0}^{2} / a_{0}^2 \ll 1$ condition, which defines the regime in which the aforementioned advantages of non-classical light are most prominent. A field that is constrained to a radius-$a_{0}$ cross-section and a spatial-bandwidth $1/\rho_{0}$ consists of approximately $a_{0}^{2}/\rho_{0}^{2}$ independent spatial modes, and the spectral mode density for a stationary field with coherence time $T_{0}$ is given by $1/T_{0}$. Hence, $P T_{0} \rho_{0}^{2} / a_{0}^2$ is the source's average photon number per spatio-temporal mode, which must be much less than one for \eqref{GSM:PS_NCL} to be an accurate approximation. However, the source has low spatial-coherence, $\rho_{0}^{2} / a_{0}^2 \ll 1$, so that the total number of photons it emits per coherence time can be large, $P T_{0} \gg 1$, while still satisfying the condition $P T_{0} \rho_{0}^{2} / a_{0}^2 \ll 1$. Therefore, the resolution and field-of-view improvements achieved with non-classical light do not require operation in the biphoton limit, wherein each signal-reference photon pair can be time resolved by ~MHz-bandwidth photon-counting detectors. 

Our analysis thus far assumed that the detector plane coincides with the object plane, but a realistic ghost-imaging scenario will likely require a separation between the two planes, as shown in Fig.~\ref{GI:detectorplane}. Here the bucket detector is along an $L_{R}$-m-long free-space path from the object that is not under our control, but the signal-arm path may be modified freely. Thus we place a focal-length $f$ lens in the signal path at a distance $d_{1}$-m from the object plane and $d_{2}$-m from the detector plane, where $1/d_{1} + 1/d_{2} = 1/f$. In addition, because the optical path lengths may be different, we introduce a post-detection electronic delay $(L_{R} - d_{1} - d_{2})/c $ to maximize the temporal coherence of the two fields. The photocurrent cross-correlation is then $C'(\rhovec_{1}) = C_{0} + \int_{\mathcal{A}_2}\! d \rhovec_{2}\, [C_{n}  |K^{(n)}_{1',2'}(\rhovec_{1}, \rhovec_{2})|^{2} + C_{p} |K^{(p)}_{1',2'}(\rhovec_{1}, \rhovec_{2})|^{2}]$, where $K^{(m)}_{1',2'}(\rhovec_{1}, \rhovec_{2})$, for $m \in \{n,p \}$, are the propagated phase-insensitive and phase-sensitive cross correlations whose magnitudes are
\begin{equation*}
\left |\frac{k_{0}M}{2\pi L_{R}}  \int d \rhovec'  e^{-i \frac{k_{0}}{2 L_{R}}(2 \rhovec_{2} \cdot \rhovec'- |\rhovec'|^2)}  K_{1,2}^{(m)}(M \rhovec_{1},\rhovec') T(\rhovec') \right|,
\end{equation*}
with $M = - d_{2}/d_{1}$ being the signal-arm magnification factor. For a sufficiently large bucket detector we can utilize Parseval's theorem to show that $C'(\rhovec_{1}) =  M^2 C(M \rhovec_1)$, where $C(\rhovec_{1})$ is given by \eqref{GI:corr}. Thus, choosing $d_{1} = d_{2} = 2f$  yields an inverted version of the object-plane ghost image. Image resolution and field of view are then determined by the phase-sensitive and phase-insensitive coherence properties of the object-plane fields, and the placement of the detectors relative to this plane only determines the signal-arm optics that are needed to obtain this object-plane ghost image.

So far we have concentrated on the image-bearing terms in \eqref{GI:corr}. However a background term, $C_{0}$, is also present and it degrades image contrast. To assess this effect, let us assume that both detectors have Gaussian shaped post-detection filters with effective integration time $T_{d}$, and that the temporal part of the sources' cross-correlation functions are given by \eqref{GSM:PIS} for the classical fields and by \eqref{GSM:PS_NCL} for the non-classical fields. Then, for classical fields---either phase-insensitive or phase-sensitive---the contrast is proportional to $1/\sqrt{1+T_{d}^{2}/4 T_{0}^2}$, which is close to unity when the fields are narrowband, i.e. when $T_{0} \gg T_{d}$. However for broadband fields, the contrast is proportional to the ratio of the detector bandwidth to the source bandwidth, $T_{0}/T_{d} \ll 1$, and the degradation is severe. On the other hand, for non-classical low-brightness fields---as occurs in the biphoton limit---the contrast is proportional to $( a_{0}^2 / P T_{0} \rho_{0}^{2})/\sqrt{1+T_{d}^{2}/2 T_{0}^2}$, so that high contrast is possible even with broadband fields.

It is relevant to point out that some features of biphoton ghost imaging which have been previously associated with the entanglement between the two source photons, are fundamentally due to the phase-sensitive cross correlation at the source, rather than the entanglement {\it per se}. So, when ghost imaging is performed with phase-sensitive light, image inversion occurs for both classical and quantum sources.  This is due to the difference in free-space propagation of phase-sensitive and phase-insensitive correlations \cite{ErkmenShapiro:CohThy}; it is not necessary for the phase-sensitive coherence to be stronger than classical.

In summary, ghost imaging with Gaussian-state sources is due to phase-sensitive and phase-insensitive cross correlations, with only the image contrast being affected by whether the light is classical or quantum. When ghost images from classical and quantum sources having identical auto-correlations are compared, the non-classical sources with much less than one photon per spatio-temporal mode offer resolution enhancement in near-field operation and field-of-view enhancement in far field operation, in addition to higher contrast in both regimes.  The large number of spatial modes implies that the total photon flux need not be time resolved by photon-counting detectors to reap these advantages.  Far-field spatial resolution and the near-field field-of-view are determined by the beam sizes at the source, so they are identical for classical and quantum sources.

This work was supported by the U. S. Army Research Office MURI Grant W911NF-05-1-0197.


\begin{thebibliography}{10}
\bibitem{classical}We will use quantum notation throughout our analysis.   Nevertheless, for field states that have proper $P$-representations, the same $C(\rhovec_1)$ results can be obtained from classical electromagnetism and shot-noise theory.
\bibitem{FOV}Field-of-view usually refers to a solid-angle region, but we will use the beam radius at a transverse plane.
\end{thebibliography}
\end{document}